\documentclass[aip,reprint]{revtex4-1} 
\usepackage[utf8]{inputenc}
\usepackage{amsfonts,amsmath,amssymb}
\usepackage{graphicx}
\usepackage{bm} 
\usepackage{mathrsfs}
\usepackage{subfigure}
\usepackage{textcomp}
\usepackage{microtype}
\usepackage{hyperref}
\usepackage[flushleft]{threeparttable}
\usepackage[per-mode=symbol]{siunitx}
\usepackage[version=3]{mhchem}

\usepackage{booktabs}
\graphicspath{{Figures/}}

\DeclareSIUnit\intensity{\watt\per\centi\meter\squared}
\DeclareSIUnit\fieldstrength{\volt\per\centi\meter}
\usepackage{xspace}

\DeclareUnicodeCharacter{2009}{\,}

\newcommand{\degree}{\ensuremath{^\circ}}%

\newlength{\figwidth}
\newlength{\figwidthwide}
\let\orgautoref\autoref
\providecommand{\Autoref}{%
  \def\equationautorefname{Equation}%
  \def\figureautorefname{Figure}%
  \def\subfigureautorefname{Figure}%
  \def\tableautorefname{Table}
  \orgautoref}
\renewcommand{\autoref}{%
  \def\equationautorefname{Eq.}%
  \def\figureautorefname{Fig.}%
  \def\subfigureautorefname{Fig.}%
  \orgautoref}

\usepackage{color}
\usepackage[normalem]{ulem}
\definecolor{darkgreen}{rgb}{0.0,0.7,0.0}

\begin{document}


\title{Structure determination of alkali trimers on helium nanodroplets through laser-induced Coulomb explosion} 

\author{Lorenz Kranabetter}
\thanks{These authors contributed equally to the work.}
\affiliation{Department of Chemistry, Aarhus University, Langelandsgade 140, DK-8000 Aarhus C, Denmark}
\author{Henrik H. Kristensen}
\thanks{These authors contributed equally to the work.}
\affiliation{Department of Physics and Astronomy, Aarhus University, Ny Munkegade 120, DK-8000 Aarhus C, Denmark}
\author{Constant A. Schouder}
\affiliation{Department of Chemistry, Aarhus University, Langelandsgade 140, DK-8000 Aarhus C, Denmark}
\affiliation{LIDYL, CNRS, CEA, Universit\'{e} Paris‐Saclay, 91191 Gif‐sur‐Yvette, France}
\author{Henrik Stapelfeldt}
\email[]{henriks@chem.au.dk}
\affiliation{Department of Chemistry, Aarhus University, Langelandsgade 140, DK-8000 Aarhus C, Denmark}

\date{\today}

\begin{abstract}
Alkali trimers, \ce{Ak3}, located on the surface of He nanodroplets are triply ionized following multiphoton absorption from an intense femtosecond laser pulse leading to fragmentation into three correlated \ce{Ak+} ions. Combining the information from three-fold covariance analysis of the emission direction of the fragment ions and from their kinetic energy distributions $P(E_{\text{kin}})$, we find that \ce{Na3}, \ce{K3}, and \ce{Rb3} have an equilateral triangular structure, corresponding to that of the lowest-lying quartet state $^{4}\mathrm{A}_{2}'$, and determine the equilibrium bond distance $R_\text{eq}$(\ce{Na_3})~=~4.65$\pm$~0.15~\AA, $R_\text{eq}$(\ce{K_3})~=~5.03$\pm$~0.18~\AA{}, and $R_\text{eq}$(\ce{Rb_3})~=~5.45$\pm$~0.22~\AA. For \ce{K3} and \ce{Rb3} these values agree well with existing theoretical calculations, while for \ce{Na3} the value is 0.2--0.3~\AA~larger than the existing theoretical results. The discrepancy is ascribed to a minor internuclear motion of \ce{Na3} during the ionization process. Also, we determine the distribution of internuclear distances $P(R)$ under the assumption of fixed bond angles. The results are compared to the square of the internuclear wave function $|\Psi(R)|^2$.

\end{abstract}

\maketitle

When molecules are multiply ionized, nowadays almost always achieved by exposure to an intense femtosecond laser pulse, they typically break apart into positively charged fragment ions. This process, termed Coulomb explosion~\cite{yatsuhashi_multiple_2018,schouder-arpc}, provides information about the structure of the molecule at the instant the laser pulse arrives because the molecular structure is imprinted on the emission directions and the kinetic energies of the ion fragments, i.e., on accessible experimental observables. As such, laser-induced Coulomb explosion has been applied in a large number of works to explore both the static structure of molecules in stationary states and the time-dependent structure of molecules undergoing intramolecular motions initiated by a femtosecond pump pulse.
Examples include identification of structural isomers~\cite{ablikim_identification_2016,burt_communication:_2018,bhattacharyya_strong-field-induced_2022}, determination of the absolute configuration of chiral molecules~\cite{pitzer_direct_2013,christensen_using_2015}, proton migration~\cite{ibrahim_tabletop_2014}, torsion~\cite{madsen_manipulating_2009,hansen_control_2012}, dissociation~\cite{stapelfeldt_wave_1995,legare_imaging_2005,bocharova_time-resolved_2011,burt_coulomb-explosion_2017,corrales_coulomb_2019,zhao_tracking_2021,unwin_x-ray_2023,walmsley_characterizing_2023} and roaming~\cite{endo_capturing_2020}.

The structure determination of a parent molecule from the measured momentum vectors of the fragment ions relies on the assumption that their interaction is given by Coulomb repulsion. While this assumption is sufficiently good for some of the examples listed above, non-Coulombic effects set a limitation on the accuracy of reconstructed structures. For instance, in the case of Coulomb explosion of an atomic dimer, such as \ce{Ar2}, non-Coulombic effects cause the one-to-one correspondence between the initial internuclear distance and the kinetic energy of the fragment ions, valid in the Coulomb approximation, to disappear~\cite{schouder_laser-induced_2020} unless all valence electrons are stripped. This was typically not the case for the many past studies employing intense visible or near-infrared laser pulses although recent results indicate that ultrashort x-ray pulses from free-electron lasers are better suited for this purpose~\cite{li_coulomb_2022}.
Here we study trimers of alkali-metal atoms in which case only three, relatively loosely bound electrons need to be removed to empty the valence shell. This appears as a favorable situation for accurate structure determination via laser-induced Coulomb explosion. The objective of our work is to explore if this is indeed the case.

\begin{figure}
\includegraphics[width=8.5cm]{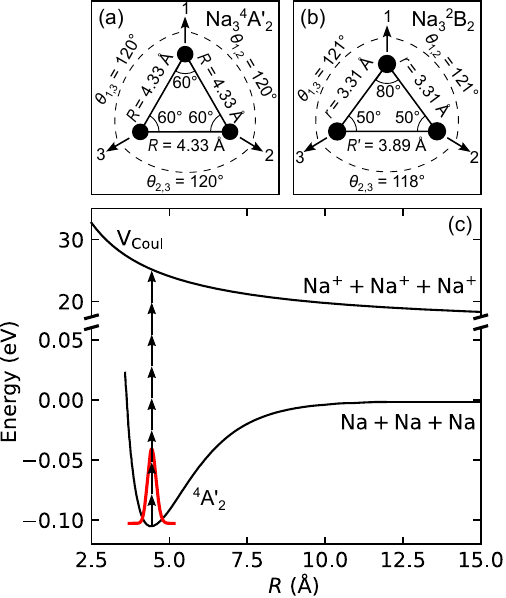}
\caption{(a),(b) Sketch of the structure of \ce{Na3} in the $^{4}\mathrm{A}_{2}'$ state~\cite{higgins_importance_2000} (a) and in the $^{2}\mathrm{B}_2$ state~\cite{thompson_analytic_1985} (b). The arrows indicate the emission angles of the \ce{Na+} fragments upon Coulomb explosion induced by triple ionization. The \ce{Na+} ion fragments are labelled 1, 2 and 3 for reference, and $\theta$ indicates the angles between the emitted fragments. (c) Potential energy curve for the $^{4}\mathrm{A}_{2}'$ state of \ce{Na3}~\cite{higgins_importance_2000} extrapolated to 15 Å and the Coulomb potential curve representing \ce{Na3^{3+}}. The red curve displays the square of the wave function of the vibrational ground state in the $^{4}\mathrm{A}_{2}'$ potential. The vertical black arrows illustrate the multiphoton triple ionization process leading to Coulomb explosion.}
\label{fig:Na3_potentials}
\end{figure}

Previous studies have shown that alkali trimers can be formed at the surface of helium nanodroplets~\cite{reho_photoinduced_2001,nagl_high-spin_2008, giese_homo-_2011}. Employing vibrationally-resolved absorption spectroscopy of electronically excited states, it was found that the trimers are mainly, if not exclusively, created in the lowest-lying quartet state, $^{4}\mathrm{A}_{2}'$. According to theory~\cite{higgins_importance_2000,hauser_relativistic_2010}, the trimers in this state have an equilateral triangular structure with $D_{\text{3h}}$ symmetry [see~\autoref{fig:Na3_potentials}(a) for \ce{Na3}]. The spectroscopic studies also showed that the trimers are populated only in the vibrational ground state, indicating that the trimers equilibrate to the 0.4~K temperature of the droplets\cite{higgins_spin_1996}. This finding is consistent with results from magnetic circular dicroism experiments~\cite{aubock_triplet_2007, aubock_observation_2008}. Despite the many experimental investigations, a direct determination of the internuclear distances of the alkali trimers has, to our knowledge, not been achieved. In the present work, we demonstrate that Coulomb explosion, triggered by an intense femtosecond pulse, identifies that \ce{Na3}, \ce{K3}, and \ce{Rb3} are formed in the $^{4}\mathrm{A}_{2}'$ state and allows determination of their equilibrium bond lengths. As such the studies here are related to previous work on Coulomb explosion imaging of trimers composed of noble gas atoms~\cite{ulrich_imaging_2011,voigtsberger_imaging_2014,xie_dynamical_2015, kunitski_observation_2015}.

Our method is based on triple ionization of alkali trimers via multiphoton absorption as illustrated in~\autoref{fig:Na3_potentials}(c) for \ce{Na3} in the $^{4}\mathrm{A}_{2}'$ state. As a result of the ionization, the vibrational wave function of \ce{Na3} is projected onto the repulsive potential curve of \ce{Na_3^{3+}}. There is only one such curve because \ce{Na_3^{3+}} has a closed-shell electron structure and this curve is well approximated by the Coulomb energy available between the three charges, $V_{\text{coul}}(R)$~=~3/$R$ in atomic units, where $R$ is the internuclear distance in the equilateral triangular structure. Upon release of this potential energy, the trimer will fragment into three \ce{Na^+} ions, each receiving a kinetic energy,
\begin{equation}
\label{eq:Coulomb}
\begin{aligned}
E_\text{kin} = V_\text{Coul}(R)/3 = 14.4~\text{eV}/R[\text{Å}].
\end{aligned}
\end{equation}
We measured $E_{\text{kin}}$ for \ce{Na3}, \ce{K3}, and \ce{Rb3} and in each case observed a peak close to the value expected for Coulomb explosion of the trimers in the $^{4}\mathrm{A}_{2}'$ state. This interpretation is supported by the relative emission directions of the fragment ions, localized at \SI{120}{\degree} and \SI{240}{\degree}, i.e., consistent with the equilateral triangular structure of trimers in the $^{4}\mathrm{A}_{2}'$ state. Furthermore, via \autoref{eq:Coulomb} we determine the distribution of internuclear separations $P(R)$ by a transformation of probabilities, using a standard Jacobian transformation.

\begin{figure}
\includegraphics[width=8.5cm]{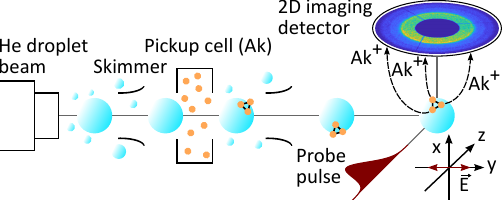}
\caption{Sketch of the core parts of the experiment. Alkali trimers \ce{Ak3} are formed on helium droplets after capture of gas-phase alkali atoms in a pickup cell. The trimers are then triply ionized and the \ce{Ak+} fragment ions recorded with a 2D imaging detector.}
\label{fig:setup}
\end{figure}

The experimental setup has been described previously~\cite{kristensen_quantum-state-sensitive_2022, kristensen_laser-induced_2023}. Thus, no elaborate details are given here. To create a continuous beam of helium droplets, a precooled gas of high purity (5.0) helium is expanded into a vacuum chamber at a stagnation pressure of 25 bars through a 5-$\mu$m-diameter nozzle. The nozzle is cooled to $T_\text{nozzle}$~=~10.5--15~K, thereby producing droplets with a mean size of $\sim$~5000--18000 He atoms~\cite{toennies_superfluid_2004}. After formation, the droplets pass through a pickup cell containing a gas of alkali atoms, see~\autoref{fig:setup}. The vapor pressure in the pickup cell is adjusted such that some of the helium droplets pick up three \ce{Ak} atoms. The three atoms can then form a trimer on the surface of the droplet~\cite{stienkemeier_laser_1995,higgins_importance_2000,nagl_heteronuclear_2008}. Hereafter, the doped droplets enter a velocity map imaging (VMI) spectrometer inside which a pulsed ($\tau_{\text{FWHM}} \approx$~50~fs) linearly polarized, focused laser beam intersects the droplet beam. The intensity and central wavelength of the probe pulses are specified in~\autoref{fig:energies}. The VMI spectrometer projects the \ce{Ak+} ions created by the laser pulses onto a 2D imaging detector monitored by a CCD camera. The detector is gated such that it only detects ions with a specific mass-to-charge ratio (\ce{^23Na+}, \ce{^39K+}, \ce{^85Rb+}, and \ce{^133Cs+}). The camera records ion hits from 10 laser shots in each frame.

\begin{figure*}
\includegraphics[width=17cm]{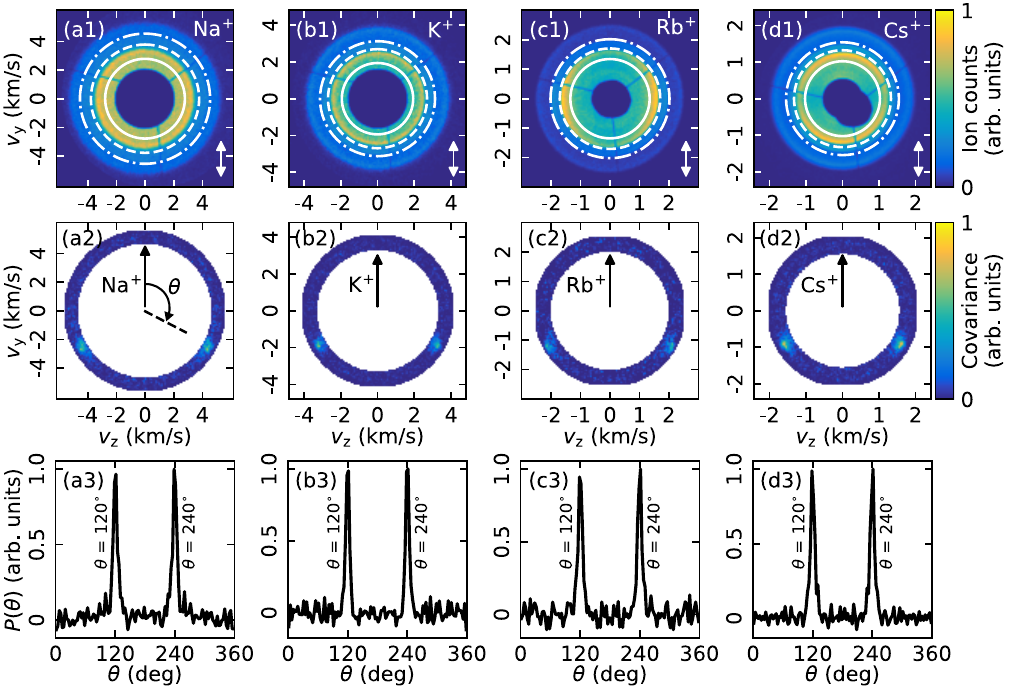}
\caption{(a1)--(d1) 2D velocity images of \ce{Ak+} ions.  The annotated white rings mark the regions pertaining to the alkali dimers and to the alkali trimers, the latter being the ions outside the dot-dashed ring, see the text. The annotated white arrows in the bottom right corners of each panel show the polarization axis of the laser pulses. For the data in (a1) and (c1) $T_{\text{Nozzle}}$ was 11~K, while for (b1) and (d1) it was 12~K. (a2)--(d2) Three-fold covariance images for the ions from \ce{Ak3} in 2D velocity map images. For improved visual contrast, negative values have been set to zero. (a3)--(d3) The angular distribution $P(\theta)$  for the covariance images in (a2)--(d2). The central positions of the two peaks are annotated next to the peaks. The FWHM of the peaks are \SI{9}{\degree} for b3) and \SI{11}{\degree} for a3), c3), and d3).}
\label{fig:ion_covariances}
\end{figure*}

\begin{figure}
\includegraphics[width=8.5cm]{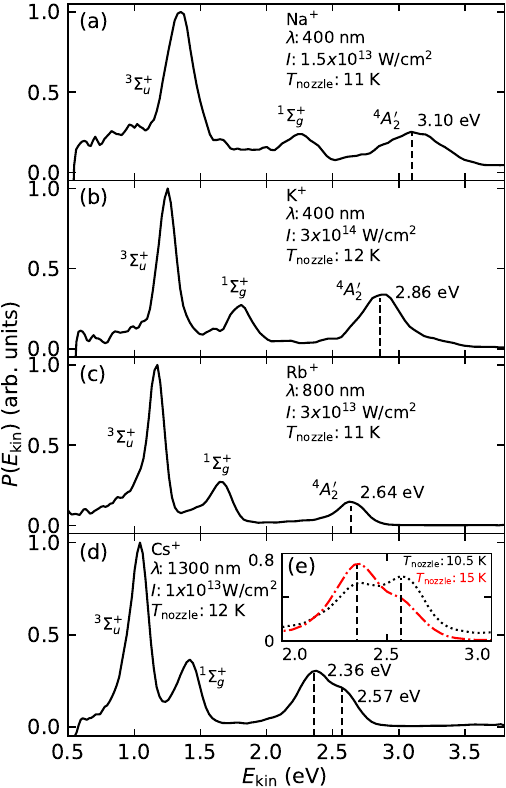}
\caption{(a)--(d) Kinetic energy spectra for \ce{Na+}, \ce{K+}, \ce{Rb+}, and \ce{Cs+} ions extracted from the 2D velocity images shown in~\autoref{fig:ion_covariances}(a1)--(d1). (e) Close-up view of the \ce{Cs+} $P(E_{\text{kin}})$ peak structure associated with trimers, measured with $T_{\text{Nozzle}}$~=~10.5~K (black dotted line) and $T_{\text{Nozzle}}$~=~15~K (red dot-dashed line, scaled by a factor of 10). The
vertical dashed lines mark the center of the peaks ascribed to trimers.}
\label{fig:energies}
\end{figure}

\Autoref{fig:ion_covariances}(a1)--(d1) presents the 2D velocity images recorded for various alkali ions. Each image consists of tens of thousands of frames stacked into one final image. Three radial stripes and a center with no signal is present in all images. The lacking signal is due to a metal disk mounted on support rods in front of the detector~\cite{schouder_laser-induced_2020, chatterley_laser-induced_2020}. The metal disk screens the detector from unwanted low-speed \ce{Ak+} ions created during ionization of droplets doped with only one \ce{Ak} atom or ionization of isolated effusive \ce{Ak} atoms present inside the VMI spectrometer. To improve the visual contrast of the outer features of the images, the centers in~\autoref{fig:ion_covariances}(a1), (b1), and (d1) have been cut digitally to remove the remaining few \ce{Ak+} ions that passed by the edge of the metal disk. Outside the screened center, three radially separated channels can be observed in the images. The innermost channel, lying between the annotated solid and dashed circles, has previously been identified as \ce{Ak+} fragments from double ionization of \ce{Ak2} in the 1$^3\Sigma_{u}^+$ state, while the channel between the dashed and dot-dashed circles were identified as \ce{Ak+} fragments from double ionization of \ce{Ak2} in the 1$^1\Sigma_{g}^+$ state~\cite{kristensen_quantum-state-sensitive_2022, kristensen_laser-induced_2023}. In this work, we focus on the last channel outside the dot-dashed rings.

To investigate the \ce{Ak+} ions in the outermost channel, we calculate the three-fold covariance~\cite{zhaunerchyk_theory_2014} image for the \ce{Ak+} ions located outside the dot-dashed rings in a 2D velocity ion image. Three-fold covariance has previously been used in gas-phase experiments to identify correlations between three ionic fragments~\cite{pickering_communication:_2016, vallance_covariance-map_2021,allum_multi-particle_2021}. A useful representation is achieved by transforming the coordinate system of the measurements into the recoil frame of one of the \ce{Ak+} fragments. This \ce{Ak+} fragment is designated the 'reference ion' and the covariance between the reference ion and the other identical \ce{Ak+} fragments in the outermost channel within the same camera frame is calculated, forming a covariance image in the recoil frame of the reference ion. This is done for all ion events in the channel. Finally, all the covariance images are rotated such that the reference ions lie along one common vector and summed together. The result is a recoil frame covariance image where the covariances with two partner ions are plotted relative to the reference ions, which now all lie along one direction (vertical). The covariance images obtained for \ce{Na3}, \ce{K3}, \ce{Rb3}, and \ce{Cs3} are displayed in~\autoref{fig:ion_covariances}(a2)--(d2).

Each covariance image shows two areas with a significantly enhanced covariance. To gain further insight, we integrate the covariance images along the radial axis and obtain the angular distribution $P(\theta)$ of the covariance, see~\autoref{fig:ion_covariances}(a3)--(d3). We then determine the central positions of the two major peaks in $P(\theta)$, which correspond to the two areas in the covariance images. The peak positions [see annotations on~\autoref{fig:ion_covariances}(a3)--(d3)] show that the emission direction of the reference \ce{Ak+} ion is correlated with two departing \ce{Ak+} ions at \SI{120}{\degree} and \SI{240}{\degree}, corresponding to an emission angle of \SI{120}{\degree} between neighbouring fragments. This is in agreement with the equilateral structure of the $^{4}\mathrm{A}_{2}'$ state trimers, as illustrated for \ce{Na3} on figure~\autoref{fig:Na3_potentials}(a). A classical trajectory calculation reveals, however, that Coulomb explosion via triple ionization of \ce{Ak3} in the ground state $^{2}\mathrm{B}_{2}$, an obtuse triangular structure with $C_{\text{2v}}$ symmetry, will eject \ce{Ak+} ions with emission angles very similar to those observed for the $^{4}\mathrm{A}_{2}'$ state. In the case of \ce{Na3} [see~\autoref{fig:Na3_potentials}(b)] and \ce{Cs3}, the emission angle between neighbouring ion fragments will be \SI{121}{\degree} and \SI{118}{\degree}, while the angles will be \SI{121}{\degree} and \SI{117}{\degree} for \ce{K3} and \ce{Rb3}. We are unable to resolve such small differences in the emission angles, due to the \SI{9}{\degree} FWHM of the peaks in~\autoref{fig:ion_covariances}(b3) and the \SI{11}{\degree} FWHM of the peaks in \autoref{fig:ion_covariances}(a3), (c3), and (d3). Thus, the three-fold covariance images are consistent with \ce{Ak3} in either the $^{2}\mathrm{B}_{2}$ or the $^{4}\mathrm{A}_{2}'$ state, but they cannot by themselves distinguish the two states.

To obtain more information about the observed triangular structures, we determine the kinetic energy distributions $P(E_{\text{kin}})$ of the \ce{Ak+} ions. For this task we employ an Abel inversion algorithm~\cite{roberts_toward_2009} to retrieve the radial velocity distributions from the ion images. These distributions are then converted to $P(E_{\text{kin}})$ by applying an energy calibration for the VMI spectrometer and then finally a standard Jacobian transformation~\cite{schouder_laser-induced_2020}. \Autoref{fig:energies} displays $P(E_{\text{kin}})$ for all \ce{Ak+}. For \ce{Na+}, \ce{K+}, and \ce{Rb+} the distribution contains three significant peaks, of which two can be assigned as \ce{Ak+} ions produced from Coulomb explosion of \ce{Ak2} in either the 1$^1\Sigma_{g}^+$ or the 1$^3\Sigma_{u}^+$ state, as demonstrated previously~\cite{kristensen_quantum-state-sensitive_2022, kristensen_laser-induced_2023}. These peaks are labeled correspondingly in~\autoref{fig:energies}. Here our interest is the peaks at higher energies, which were not identified before.

For \ce{Na+}, see~\autoref{fig:energies}(a), the unidentified peak is centered at 3.10~eV (determined by a Gaussian fit). Coulomb explosion of \ce{Na3} in the $^{4}\mathrm{A}_{2}'$ state at the equilibrium distance, $R_\text{eq}$~=~4.41~\AA~\cite{higgins_importance_2000} would produce three \ce{Na+} fragments with $E_{\text{kin}}$ = 3.3~eV. In comparison, Coulomb explosion of \ce{Na3} in the $^{2}\mathrm{B}_{2}$ state, would result in one \ce{Na+} ion with $E_{\text{kin}}$~=~4.3~eV [fragment 1, see~\autoref{fig:Na3_potentials}(b)] and two \ce{Na+} ions with $E_{\text{kin}}$~=~4.0~eV (fragment 2 and 3). Therefore, we interpret the peak at 3.10~eV as ions formed by Coulomb explosion of \ce{Na3} in the $^{4}\mathrm{A}_{2}'$ state. Similarly, the center of the high-energy peaks, marked by vertical dashed lines, for the \ce{K+} (2.86~eV) and \ce{Rb+} (2.64~eV) ions in \autoref{fig:energies}(b) and (c) match well with the kinetic energy each fragment ion receives upon Coulomb explosion of trimers in the $^{4}\mathrm{A}_{2}'$ state and, importantly, do not match the energy of 3.5 and 3.2~eV (3.2 and 3.0~eV) fragment ions would get from Coulomb explosion of \ce{K3} (\ce{Rb3}) in the $^{2}\mathrm{B}_{2}$ state. As such, we also conclude that \ce{K3} and \ce{Rb3} are produced in the $^4\mathrm{A}_{2}'$ state~\cite{double_ionization}.

From the central position of the peaks in $P(E_{\text{kin}})$, we determine the equilibrium internuclear distance $R_{\text{eq}}$ for the trimers in the identified $^{4}\mathrm{A}_{2}'$ state via \autoref{eq:Coulomb}. \Autoref{tab:comparison} provides a summary of the identified peaks and the corresponding equilibrium bond distances, alongside theoretical values for $R_{\text{eq}}$ previously reported in the literature. The uncertainty on the peak positions in $P(E_{\text{kin}})$ results from a combination of the energy resolution of the VMI spectrometer, the slightly asymmetric shape of the peaks and, for \ce{K3} and \ce{Rb3}, the presence of different isotopologues~\cite{isotopologue}. For \ce{K3} and \ce{Rb3} the determined $R_{\text{eq}}$ is within 0.1~Å of the theoretical values, while for \ce{Na3} the measured $R_{\text{eq}}$ is 0.2--0.3 Å larger than the theoretical values. The reason for this discrepancy is, we believe, a minor internuclear motion on potential curves in \ce{Na3}, \ce{Na3^+} or in \ce{Na3^{2+}} transiently excited during the laser pulse. Its 50 fs duration is long enough that such internuclear motion can occur at least for \ce{Na3}, which is the lightest of the trimers. A similar shift of the equilibrium internuclear distance was also observed for Coulomb explosion studies of \ce{Na2} on helium droplets whereas $R_{\text{eq}}$ measured for \ce{K2} and \ce{Rb2} were very close to the theoretical values~\cite{kristensen_laser-induced_2023}.

\begin{table}
\begin{ruledtabular}
\centering
\caption{Central position of the peak in $P(E_{\text{kin}})$ pertaining to Coulomb explosion of \ce{Ak3} in the $^{4}\mathrm{A}_{2}'$ state. Theoretical values from the literature are provided for $R_{\text{eq}}$.}
\label{tab:comparison}
\begin{tabular}{@{}lllll}
      & \multicolumn{2}{l}{$P(E_{\text{kin}})$ peak (eV)} & \multicolumn{2}{l}{$R_{\text{eq}}$ (Å)} \\ \midrule

\ce{Na_3} & \multicolumn{2}{l}{3.10~$\pm$~0.10}  & \multicolumn{2}{l}{4.65~$\pm$~0.15}    				\\
\ce{Na_3} Theory & \multicolumn{2}{l}{--} & \multicolumn{2}{l}{4.33\footnotemark[1], 4.41\footnotemark[2]}	\\

\ce{K_3}  & \multicolumn{2}{l}{2.86~$\pm$~0.10}  & \multicolumn{2}{l}{5.03~$\pm$~0.18}	  				\\
\ce{K_3} Theory & \multicolumn{2}{l}{--} & \multicolumn{2}{l}{4.95\footnotemark[1], 5.05\footnotemark[3]}	\\

\ce{Rb_3} & \multicolumn{2}{l}{2.64~$\pm$~0.10}  & \multicolumn{2}{l}{5.45~$\pm$~0.22}  			    \\
\ce{Rb_3} Theory & \multicolumn{2}{l}{--} & \multicolumn{2}{l}{5.35\footnotemark[1], 5.50\footnotemark[3], 5.31\footnotemark[4], 5.45\footnotemark[5]}	\\
	
\end{tabular}
\end{ruledtabular}
\footnotetext[1]{From Ref.~\onlinecite{soldan_lowest_2010}.}
\footnotetext[2]{From Ref.~\onlinecite{higgins_importance_2000}.}
\footnotetext[3]{From Ref.~\onlinecite{hauser_relativistic_2010}.}
\footnotetext[4]{From Ref.~\onlinecite{schnabel_towards_2021}.}
\footnotetext[5]{From Ref.~\onlinecite{soldan_potential_2010}.}
\end{table}

We also conducted the experiment with cesium. In this case, an unidentified broad structure composed of two individual peaks is observed in $P(E_{\text{kin}})$ with peaks at 2.36~eV and 2.57~eV (determined by Gaussian fits), see~\autoref{fig:energies}(d). Three-fold covariance analysis of the associated ions in the VMI image, see~\autoref{fig:ion_covariances}(d1)--(d3), shows that these ions originate from \ce{Cs3}. To understand the origin of the two peaks, we varied the mean droplet size by adjusting the nozzle temperature to $T_\text{nozzle}$~=~10.5~K and 15~K. The measured kinetic energy distributions are depicted in the inset in ~\autoref{fig:energies}(d), with the distribution for $T_\text{nozzle}$~=~15~K scaled by a factor of 10. We observed that the peak at 2.36~eV was dominating for the 15~K measurement, which indicate that those ions should come from Coulomb exploded \ce{Cs3} in the high-spin state $^{4}\mathrm{A}_{2}'$, as the high-spin trimers are better preserved than low-spin trimers on helium droplets, especially on small ones. The released energy of 2.36~eV is, however, a little lower than the expected kinetic energy release $E_{\text{kin}}$~=~2.59~eV for Coulomb explosion of \ce{Cs3} in the $^{4}\mathrm{A}_{2}'$ state starting from the equilibrium distance $R_{\text{eq}}$ = 5.56~Å~\cite{soldan_lowest_2010}. In fact, the expected energy release matches the other peak observed at 2.57~eV. For the 10.5~K measurement we observe that the peak at 2.57~eV begin to dominate, which is unexpected since the peak containing ions from Coulomb explosion of the high-spin state, whichever of the two peaks that is, should dominate at all droplet sizes. Finally, Coulomb explosion of \ce{Cs3} at the equilibrium distance of the $^{2}\mathrm{B}_{2}$ state~\cite{guerout_core_2009} would produce \ce{Cs+} fragments with energies of 3.0 and 2.8~eV, which does not match the observed peaks. Thus, the observations do not currently enable us to unambiguously determine the quantum states for \ce{Cs3}.

\begin{figure}
\includegraphics[width=8.5cm]{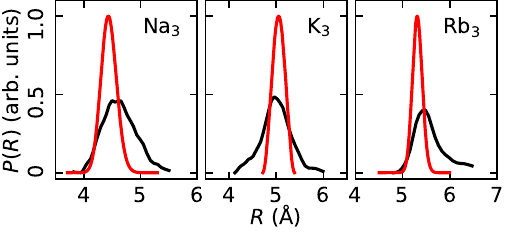}
\caption{$P(R)$ distributions for \ce{Na3}, \ce{K3}, and \ce{Rb3} in the $^{4}\mathrm{A}_{2}'$ state, determined by transformation of the measured $P(E_{\text{kin}})$ and a flat background subtraction. The red lines show the theoretical $|\Psi(R)|^2$ distributions for the isolated $^{4}\mathrm{A}_{2}'$ state.}
\label{fig:WF}
\end{figure}

Inspired by recent results on laser-induced Coulomb explosion of alkali dimers~\cite{kristensen_laser-induced_2023}, we now discuss the possibility of determining the distribution of internuclear distances $P(R)$ from $P(E_{\text{kin}})$. In the dimer case, $R$ is the only internuclear (vibrational) coordinate. Therefore, it is straightforward to determine $P(R)$ from $P(E_{\text{kin}})$ via $E_\text{kin} = 7.2~\text{eV}/R[\text{\AA}]$, the equivalent of \autoref{eq:Coulomb}, using the appropriate transformation of probability distributions. In the trimer case, the situation is more complex because there are three vibrational degrees of freedom, i.e., the pairwise distances between the three nuclei, $R_1$, $R_2$, and $R_3$ are not necessarily the same. As such, the internuclear wave function depends on three variables and can, for instance, be expressed as $\Psi(R_1,R_2,R_3)$. Measurement of the kinetic energy of the \ce{Ak+} fragment ions does not immediately enable determination of $|\Psi(R_1,R_2,R_3)|^2$ as a function of three variables. Therefore, we make the simplifying assumption that the trimer retains its equilateral triangular shape, i.e., that it is only the symmetric-stretch vibration that is important for the zero-point vibrations. This assumption, which was adopted in previous work on laser-induced Coulomb explosion of neon and argon trimers~\cite{ulrich_imaging_2011}, means that the wave function can be written as $\Psi(R)$. Thus, we can apply the procedure from the dimer case to determine $P(R)$ using \autoref{eq:Coulomb} to connect $E_{\text{kin}}$ and $R$.

The results for \ce{Na3}, \ce{K3}, and \ce{Rb3} are depicted in \autoref{fig:WF} together with $|\Psi(R)|^2$ calculated by solving the stationary vibrational one-dimensional Schrödinger equation for isolated \ce{Na3}, \ce{K3}, and \ce{Rb3} in the $^{4}\mathrm{A}_{2}'$ state using potentials from the literature~\cite{higgins_importance_2000, schnabel_towards_2021, hauser_private}. The FWHM of the theoretical $|\Psi(R)|^2$ are 0.3~Å for \ce{Na3}, 0.3~Å for \ce{K3}, and 0.2~Å for \ce{Rb3} and the FWHM of the measured $P(R)$ are about 2--3 times larger. This is very similar to that observed for the alkali dimers and we believe the factors causing the larger widths of $P(R)$ compared to those of $|\Psi(R)|^2$ are the same as for the dimer case, i.e., in particular the energy resolution of the VMI spectrometer and the distortion of the trajectories of the recoiling \ce{Ak+} ions due to their interaction with the droplet surface~\cite{kristensen_laser-induced_2023}.

In summary, we showed that Coulomb explosion of alkali trimers, triggered by laser-induced triple ionization, produces three \ce{Ak+} fragment ions that recoil with a pairwise angle of \SI{120}{\degree}. While this observation is consistent with trimers in both the $^{2}\mathrm{B}_{2}$ and the $^4\mathrm{A}_{2}'$ state, the kinetic energy of the fragment ions unambiguously show that \ce{Na3}, \ce{K3}, and \ce{Rb3} are populated in the $^4\mathrm{A}_{2}'$ state where the structure is that of an equilateral triangle. The quantum-state-sensitivity through measurement of $E_\text{kin}$ of the fragment ions is similar to that demonstrated for alkali dimers where the distinction is between the 1$^3\Sigma_{u}^+$ state and the 1$^1\Sigma_{g}^+$ state~\cite{kristensen_quantum-state-sensitive_2022,albrechtsen_hetero_2023}. Furthermore, the measured $E_\text{kin}$ of the \ce{Ak+} ions enabled the first experimental determination of the equilibrium bond distance for \ce{Na3}, \ce{K3}, and \ce{Rb3} in the $^4\mathrm{A}_{2}'$ state.
We expect that the accuracy of the bond distances can be improved by employing a VMI spectrometer with a higher energy resolution and, for \ce{K3} and \ce{Rb3}, by selecting individual isotopologues through coincident filtering of the recoil ions~\cite{albrechtsen_hetero_2023}. One future application of the results presented here, is to excite vibrational wave packets in the alkali trimers with a femtosecond pump pulse and follow their evolution with timed Coulomb explosion. For instance, it should be possible to excite the symmetric-stretch vibration in the $^4\mathrm{A}_{2}'$ state by stimulated Raman transitions~\cite{claas_wave_2006,madsen_combined_2009,shu_femtochemistry_2017}. Tracking the internuclear distance with timed Coulomb explosion could be an informative way to explore if the vibration stays localized in the initial mode or if it couples to other modes through intramolecular vibrational redistribution.

\begin{acknowledgments}

We acknowledge valuable discussions with James Pickering. We also thank Jan Thøgersen for carefully maintaining the laser system. H.S. acknowledges support from The Villum Foundation through Villum Investigator Grant No. 25886.
\end{acknowledgments}

%
%


\begin{thebibliography}{56}
\bibitem [{dou()}]{double_ionization}%
  \BibitemOpen
  \href@noop {} {\bibinfo  {journal} {If the probe pulse only doubly ionizes
  the trimers, fragmentation of \ce{Ak3^{2+}} can occur as \ce{Ak3^{2+}}
  $\rightarrow$ \ce{Ak2^{+}} + \ce{Ak+} or as \ce{Ak3^{2+}} $\rightarrow$
  \ce{Ak+} + \ce{Ak+} + \ce{Ak}. We have not identified any peaks in the
  kinetic energy distributions at the energies expected for these channels,
  which are significantly less than the energies of the \ce{Ak+} ions from
  break-up of \ce{Ak3^{3+}}. Therefore, we conclude that the alkali trimers are
  mainly undergoing triple ionization at the parameters of the laser pulses
  used}\ }\BibitemShut {NoStop}%
\bibitem [{iso()}]{isotopologue}%
  \BibitemOpen
\bibfield  {journal} {  }\href@noop {} {\bibinfo  {journal} {The effect is most
  pronounced for \ce{K3}, where there are two main isotopologues, \ce{(^39K)3}
  (81.2~\% abundance) and \ce{^39K^39K^41K} (17.5~\% abundance). For the
  latter, \autoref{eq:Coulomb} is replaced by \begin{equation*} E_\text{kin} =
  \frac{3}{2 + 39/41}\frac{14.4~\text{eV}}{R[\text{Å}]}.\end{equation*} Thus,
  \ce{^39K+} ions from Coulomb explosion of \ce{^39K^39K^41K} emerges with 50
  meV energy more than \ce{^39K+} ions from Coulomb explosion of \ce{(^39K)3}}\
  }\BibitemShut {NoStop}%
\end{thebibliography}

%

\end{document}